\documentclass[amsmath,amssymb,aps,prl,twocolumn]{revtex4-1}

\usepackage{amssymb}
\usepackage{natbib}
\usepackage{graphicx}
\usepackage{amsmath}
\usepackage{upgreek}
\usepackage{physics}
\usepackage{wasysym}
\usepackage[bookmarks = false]{hyperref}
\usepackage{color}

\begin{document}

\title{Cavity-Enhanced Atom-Photon Entanglement with Subsecond Lifetime}

\author{Xu-Jie Wang$^{1,\,2,\,*}$}
\author{Sheng-Jun Yang$^{1,\,2,\,*}$}
\author{Peng-Fei Sun$^{1,\,2}$}
\author{Bo Jing$^{1,\,2}$}
\author{Jun Li$^{1,\,2}$}
\author{Ming-Ti Zhou$^{1,\,2}$}
\author{Xiao-Hui Bao$^{1,\,2}$}
\author{Jian-Wei Pan$^{1,\,2}$}

\affiliation{$^1$Hefei National Laboratory for Physical Sciences at Microscale and Department
of Modern Physics, University of Science and Technology of China, Hefei,
Anhui 230026, China}
\affiliation{$^2$CAS Center for Excellence in Quantum Information and Quantum Physics, University of Science and Technology of China, Hefei, Anhui 230026, China}
\affiliation{$^*$These two authors contributed equally to this work.}

\begin{abstract}
A cold atomic ensemble suits well for optical quantum memories, and its entanglement with a single photon forms the building block for quantum networks that give promise for many revolutionary applications. Efficiency and lifetime are among the most important figures of merit for a memory. In this paper, we report the realization of entanglement between an atomic ensemble and a single-photon with subsecond lifetime and high efficiency. We engineer dual control modes in a ring cavity to create entanglement and make use of 3-dimensional optical lattice to prolong memory lifetime. The memory efficiency is 38\% for 0.1~second storage. We verify the atom-photon entanglement after 1~second storage by testing the Bell inequality with a result of $S=2.36\pm0.14$.
\end{abstract}

\maketitle

A quantum network~\cite{Kimble2008,Wehner2018} of long-lived quantum memories~\cite{Lvovsky2009,Bussieres2013,Afzelius2015,Heshami2016} gives the promise for a series of revolutionary applications~\cite{Wehner2018}, such as large-scale quantum communication via quantum repeaters~\cite{Briegel1998}, cooperative operation of atomic clocks~\cite{Komar2014}, and distributed quantum sensing. The building block for quantum networks is a pair of hybrid entanglement between a single photon and a quantum memory~\cite{Kimble2008,Afzelius2015}, with the memory being efficient and long-lived~\cite{Razavi2009,Sangouard2011,Simon2017}. Entangling remote quantum memories is further mediated via photons through entanglement swapping~\cite{Zukowski1993,Pan1998,Simon2003}. Larger networks can be further constructed via concatenating entanglement swapping of the memories~\cite{Briegel1998,Wallnofer2016}. Hybrid memory-photon entanglement has been created in many physical systems, such as trapped ions~\cite{Duan2010}, single neutral atoms~\cite{Reiserer2015,Volz2006}, atomic ensembles~\cite{Sangouard2011}, rare earth ions~\cite{Clausen2011,Saglamyurek2011}, NV centers~\cite{Gao2015} and quantum dots~\cite{Gao2015}. Within all demonstrations so far, Dudin et~al achieved a memory-photon entanglement~\cite{Dudin2010} with the longest lifetime. While the memory efficiency in \cite{Dudin2010} was merely 7\% at 0.1~s, which hinders further scalable extensions.

Atomic ensembles~\cite{Duan2001,Hammerer2010,Sangouard2011} are an excellent candidate for quantum networks, since single photons can interact with the atoms strongly via collective enhancement. Extensive studies~\cite{Sangouard2011,Bao2012,Xu2013,Yang2015,ding2015,Rui2015,Jiang2016,Yang2016,Tian2017,Pu2017,Shi2017,Hsiao2018,vernaz-gris2018,Wang2019,Chang2019,Li2019,cao2020efficient} have been performed so far for improving the memory performances. Motional dephasing can be tackled with optical lattice~\cite{zhao2009long,Yang2016} and spin-wave engineering~\cite{Zhao2009,Bao2012,Rui2015,Jiang2016}. High efficiency can be achieved by engineering either a huge optical depth~\cite{Hsiao2018,vernaz-gris2018,Wang2019,cao2020efficient} or a low-finesse cavity with moderate optical depths~\cite{Simon2007,Bao2012,Yang2015,Yang2016}. In a previous work~\cite{Yang2016}, we have taken a joint approach with optical lattice and a ring cavity to achieve efficient and long-lived storage simultaneously. Nevertheless, we merely demonstrated nonclassical correlation of DLCZ storage~\cite{Duan2001}. In this paper, we report the realization of atom-photon entanglement with long lifetime and high retrieval efficiency. We make use of a scheme with dual control modes to create entanglement between a single photon's polarization and the momentum vector of a collective atomic excitation. Moreover, we optimize compensation of differential light shifts of the optical lattice, and get a 1/e memory lifetime of 458(19)~ms for qubit storage. The initial retrieval efficiency measured is 58\% and drops to 38\% at 0.1~s, which is 5.4 times higher than the result~\cite{Dudin2010} by Dudin et~al. Finally we test the atom-photon entanglement after 1~s storage via violation of the Bell inequality by 2.57 standard deviations ($S=2.36\pm0.14$).

\begin{figure*}[htb]
\includegraphics[width=1.4\columnwidth]{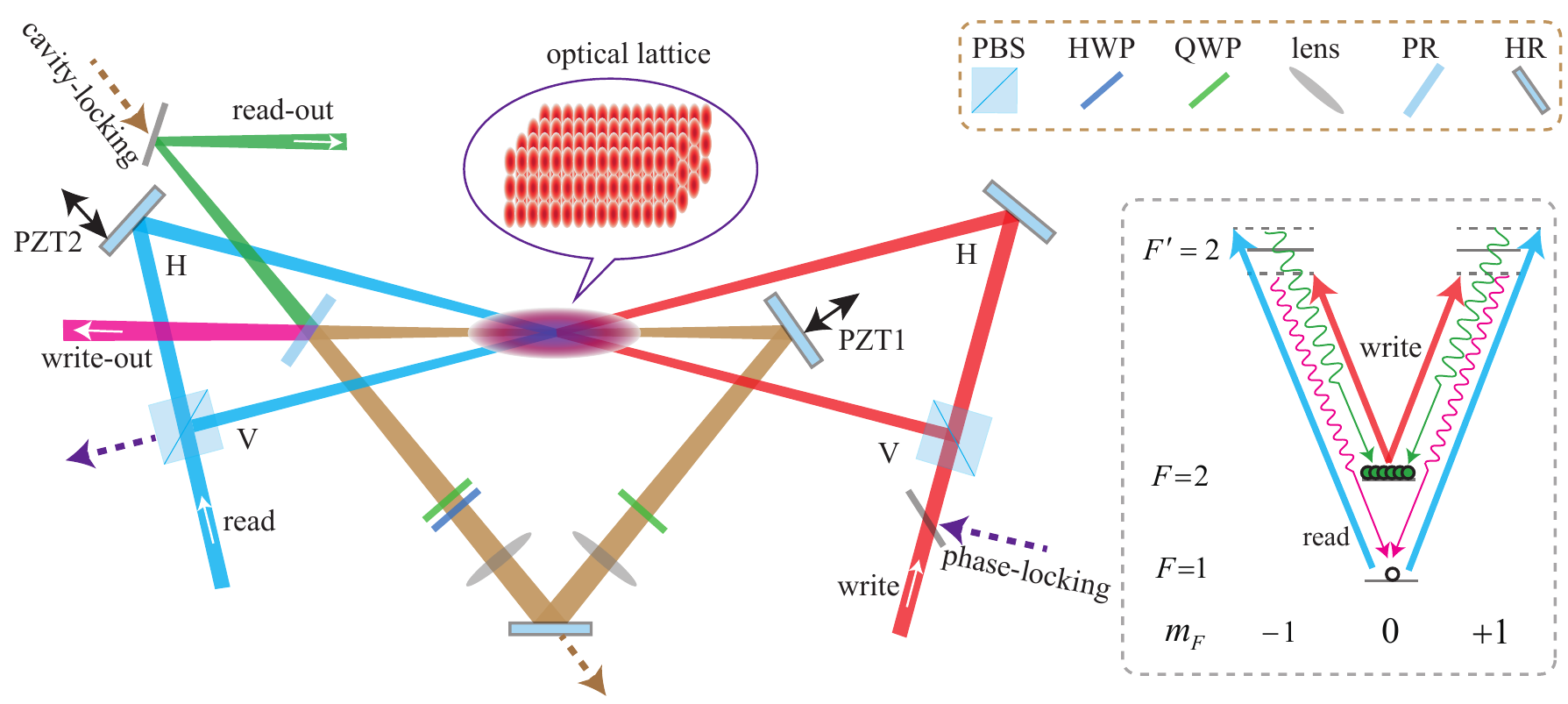}
\caption{Schematic experimental setup and atomic energy levels. The $^{87}\!\rm{Rb}$ atoms are trapped in a 3D optical lattice formed by interfering four circularly polarized 1064\,nm laser beams. The lattice beams are not shown in the picture for the sake of simplicity, and more details about the optical lattice can be found in our former publication~\cite{Yang2016}. The phase difference between two write-read paths is locked by a PID control loop. Angles between the write-out signal and the two write beams are $2.5^\circ$ and $-4.2^\circ$ respectively. A three-mirror cavity is placed around the atomic ensemble to increase the retrieval efficiency with whose beam waist matching  the write(read)-out signal. The cavity has a beam waist radius of $\sim\!60\,\rm{\upmu m}$, a finesse of 43.4/44.3 for H/V polarized light, and a free spectral range of 484 MHz. The cavity is also locked by a PID control loop. PZT1 (piezoelectric ceramic transducer) is used to lock the cavity by displacing a HR mirror, and PZT2 is used to lock the W-R interferometer. Some other parameters and symbols are listed as following. Cavity-locking beam: 796 nm, $0.5\;\upmu W$. Phase-locking beam: 796 nm, $0.5\; \upmu W$. Write beam: red detuned $-30$ MHz, $2\; \upmu W$. Read beam: blue detuned $+36$ MHz, $320\;\upmu W$. Beam radii of the write and read lasers are both 300 $\upmu$m. PBS: polarizing beam splitter. HWP: half-wave plate. QWP: quarter-wave plate. PR: partially reflecting mirror with reflectivity of 90.4(2)\%. HR: highly reflecting mirror. Lens: focal length 250 mm.}
\label{fig:Setup}
\end{figure*}

Our experimental setup is shown in Fig.~\ref{fig:Setup}. An ensemble of $^{87}$Rb atoms is first prepared via magneto-optical trapping and further transferred into a three dimensional optical lattice. Atoms in the lattice have a population lifetime of 2$\sim$3 seconds, thus providing a solid basis for long-lived storage. Nevertheless, minor differential light shifts between the two energy levels used for storage will dephase a collective atomic excitation, resulting in a typical lifetime of several milliseconds~\cite{zhao2009long}. In our experiment, they are compensated via tuning the magnetic field and making the lattice beams circularly polarized~\cite{Yang2016}. In this way, the vector part of the differential light shift cancels the scalar part~\cite{lundblad2010}. A ring cavity with a finesse of 43.4/44.3 (for H/V polarized light) is placed around the atoms to enhance the atom-light interaction. More details on our lattice and cavity setup can be found in our previous publication~\cite{Yang2016}.

To create entanglement in our setup, a popular scheme using dual internal states~\cite{Matsukevich2005} is not suitable since it hinders long-lived storage via being magnetic-field sensitive~\cite{Dudin2010d}\cite{Dudin2013}. While another popular scheme using dual spacial modes~\cite{Chen2007} is not suitable either since it requires duplicating the cavity setup which will make the whole system over complicated. Here, instead of collecting two spatial modes for the single photons as in \cite{Chen2007} to create entanglement, we harness dual spatial modes both for the read and write beams, which involves merely minor modifications of our previous setup with optical lattice and a ring cavity~\cite{Yang2016}.

The energy levels employed are shown in the inset of Fig.~\ref{fig:Setup}. Atoms are first optically pumped to the initial state $5S_{1/2}:\!|F\!=\!2,m_F\!=\!0\rangle$. Then we apply a write pulse with linear polarization to induce spontaneous Raman scattering. With a very small probability we will detect a scattered single photons in one spatial mode. The polarization of the scattered photon is orthogonal to the polarization of the write pulse, due to the specific dipole matrix elements involved. As shown in Fig.~\ref{fig:Setup}, a write pulse is split into two spatial modes, with one mode from top right with horizontal (H) polarization, and the other from bottom right with vertical (V) polarization. The top right mode will create a V photon along with a collective atomic excitation of $\ket{\downarrow}$, while the bottom right mode will create a H photon along with a collective atomic excitation of $\ket{\uparrow}$. Thus superposition of the two cases will create an entangled state of
\begin{equation}
\ket{\Psi}=\frac{1}{\sqrt{2}}(\ket{H}_w\ket{\uparrow}_a + e^{-i\phi_w}\ket{V}_w\ket{\downarrow}_a),
\label{state1}
\end{equation}
where the subscript $w$ denotes the write-out photon, $a$ denotes the atomic ensemble, and $\phi_w$ is the phase difference between two write pulses at the location of the atoms. The write process is repeated until a write-out photon is detected, giving an production rate of $\sim800$~Hz for the atom-photon entanglement without considering the atom loading time (see Supplemental Material for detailed time sequences). Afterwards, we store the atomic qubit for a programmable duration $t$, and detect the atomic qubit by applying the read pulse. Similar to the write process, a read pulse is split into two spatial modes, one mode from top left with H polarization will convert the atomic state $\ket{\downarrow}_a$ to a read-out photon with V polarization, while the other mode from bottom left with V polarization will convert $\ket{\uparrow}_a$ to a read-out photon with H polarization. Thus equivalently, the atom-photon entanglement $\ket{\Psi}$ is converted to a photon-pair entanglement of
\begin{equation}
\ket{\Psi}_t=\frac{1}{\sqrt{2}}(\ket{H}_w\ket{V}_r + e^{-i(\phi_w+\phi_r)}\ket{V}_w\ket{H}_r),
\label{state1}
\end{equation}
where the subscript $r$ denotes the read-out photon, and $\phi_r$ is the phase difference between two read pulses at the location of the atoms. In our experiment, both $\phi_w$ and $\phi_r$ drifts slightly as a function of time, thus we actively stabilize the sum of $\phi_w$ and $\phi_r$ to 0 via inserting a phase-locking beam (796~nm). The ring cavity is engineered to be resonant both for the write-out photon and the read-out photon. The cavity is actively stabilized by inserting the cavity-locking beam (796~nm) in the read-out mode. To minimize the influence on the quantum memory, we merely stabilize the write-read interferometer and the ring cavity during the atom loading phase.
Birefringence of optical elements in the ring cavity is compensated with several low-loss wave plates\cite{Yang2015}.

We note that in the case of free space~\cite{Liao2014,ding2015} the above scheme of dual control modes is imperfect since the two atomic states $\ket{\uparrow}_a$ and $\ket{\downarrow}_a$ can be retrieved by either one of the two read modes, thus leading to a reduced retrieval efficiency by one half. In our setup, the ring cavity enhances retrieval into the cavity mode, and retrieval into other directions will be suppressed. Thus ideally the collective excitation $\ket{\uparrow}_a$ will only be retrieved by the top left mode, and $\ket{\downarrow}_a$ will only be retrieved by the bottom left mode.

\begin{figure}[htb]
\includegraphics[width=1\columnwidth]{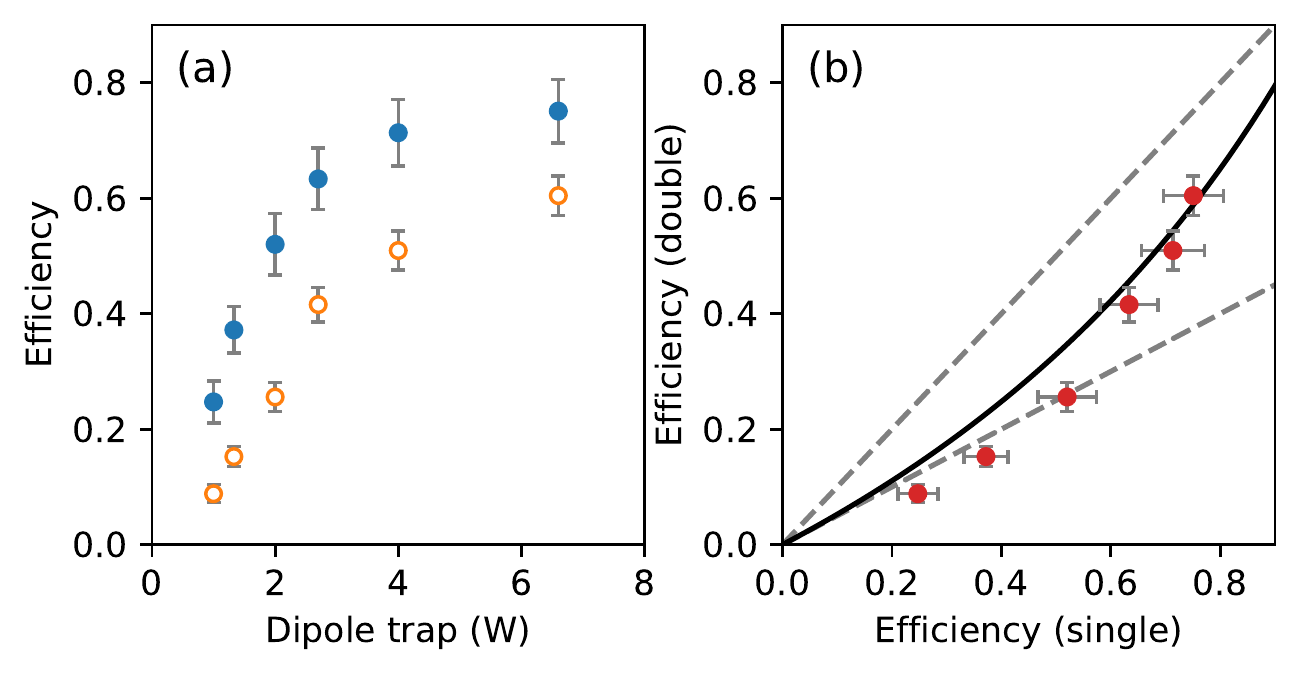}
\caption{Comparison of one and two control mode retrieval efficiencies. (a) Intrinsic retrieval  efficiencies at different lattice trap powers. Blue circles indicate memory efficiencies with a single control mode, and orange open circles indicate memory efficiencies with double control modes. (b) Double-mode efficiencies plotted as a function of single-mode efficiencies. Slopes of two dashed lines are 0.5 and 1 respectively. A solid curve in black refers to the theoretical model which is discussed in detail in Supplemental Material.}
\label{fig:efficiency}
\end{figure}

Therefore we first measure the retrieval efficiencies and compare the cases with one or two control modes, by varying the lattice trap power. We get the retrieval efficiencies via measuring the detection probabilities of a read-out photon conditioned on the detection of a write-out photon and applying loss calibration. The comparison of intrinsic retrieval efficiencies\cite{Yang2015,Yang2016} measured with dual control modes or a single control mode is shown in Fig.~\ref{fig:efficiency}(a). By increasing the trap power, we capture more atoms from the magneto-optical trap and prepare an atomic ensemble with a larger optical depth. Therefore, we see the retrieval efficiency increases as a function of trap power. To explicitly compare the two cases, we also plot the double-mode efficiency as a function of single-mode efficiency, shown in Fig.~\ref{fig:efficiency}(b). It is clear to see that, for smaller lattice trap power, the ratio of double-mode efficiency over single-mode efficiency is around 0.5. While for larger trap power, the ratio starts to rise significantly above 0.5, which is a clear signature of benefit from cavity enhancement. Extended discussions on the relation between single-mode and double-mode efficiencies are given in the Supplemental Material.

\begin{figure}[htb]
	\includegraphics[width=1\columnwidth]{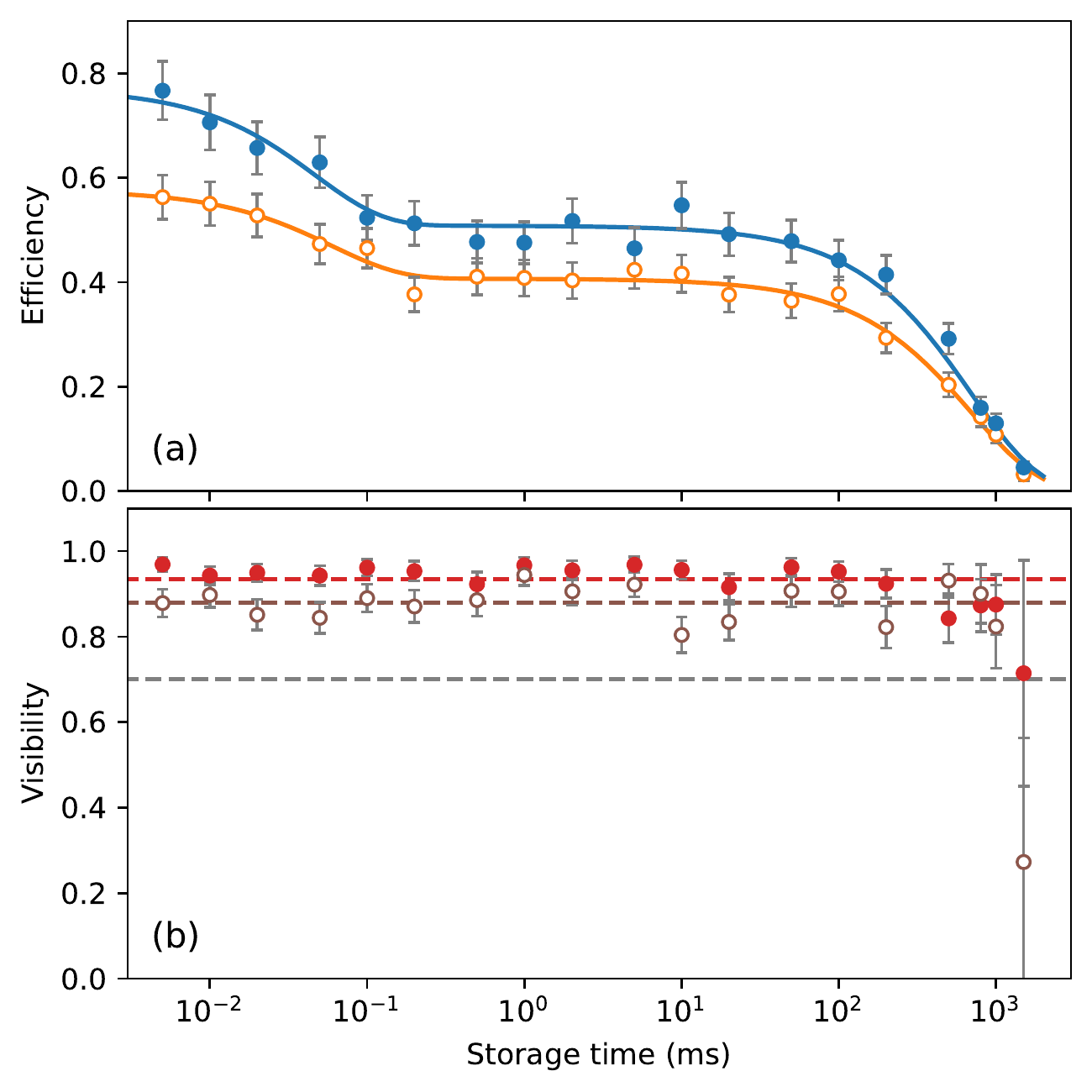}
	\caption{Intrinsic retrieval efficiencies and entanglement visibilities versus storage time. (a) Orange circles represent efficiencies of a single-mode quantum memory at different storage time. We fit the data with the function  $\eta(t)=A_1\exp(-t/\tau_1)+A_2\exp(-t/\tau_2)$, resulting in $A_1=0.26\pm 0.04$, $A_2=0.51\pm 0.01$, $\tau_1=0.047\pm 0.018$ ms, $\tau_2=697\pm 32$ ms. Blue circles represent efficiencies of the dual-mode storage. With the same fit function, we get the result of $A_1=0.17\pm 0.02$, $A_2=0.41\pm 0.01$, $\tau_1=0.060\pm 0.021$ ms, $\tau_2=703\pm 31$ ms. (b) Red solid circles show the H/V (horizontal/vertical) visibilities of the entanglement and brown open circles show the D/A (diagonal/anti-diagonal) visibilities. Red and brown dashed lines are the average values within 1 second. The grey dashed line is the threshold of 0.707 to violate the Bell inequality.}
	\label{fig:lifetime}
\end{figure}

To achieve long-lived storage, it is crucial to minimize the part of free moving atoms~\cite{Yang2016}. In our experiment, this is achieved via dynamical loading of the optical lattice. In particular, when the unconfined atoms fall off the optical lattice region, we gradually increase the trap potential within 60~ms to reduce atoms which are not totally confined within single lattice sites. The trapping potential is increased from 70~$\upmu$K to 90~$\upmu$K. Besides, it is also very important to eliminate leakage of the control beams. We find that extinction ratio of ordinary acousto-optic modulators is not high enough. Therefore, we use mechanical shutters to block all of the laser beams excluding the lattice beams during storage. Moreover, for seconds regime storage, when we apply the read pulse, the cavity resonant point and write-read interferometer locking point may both drift away. To solve this problem, for storage time larger than 100 ms, we turn on the locking beams for 3 ms a few milliseconds before the read pulse to pull the cavity back to resonance and the interferometer to the right phase. Finally, we also optimize the magnetic field gradient and magnitude very carefully. We note that during lifetime optimization, we make use of classical electromagnetically induced transparency (EIT) storage instead, which saves us quite a lot of time.

Next we measure the temporal dependence of retrieval efficiency, with the result shown in Fig.~\ref{fig:lifetime}(a). We use a double-exponential decay function $\eta(t)=A_1\exp(-t/\tau_1)+A_2\exp(-t/\tau_2)$ to fit the results. For the case of single control mode, we get an initial retrieval efficiency of 0.77(4), and 1/e lifetime of 407(42) ms. For the case of dual control modes, we get an initial retrieval efficiency of 0.58(2), 1/e lifetime of 458(35) ms. In comparison with previous single-photon storage experiments~\cite{Wang2019,cao2020efficient} involving dual modes and a huge optical depth, the lifetime achieved is more than 4 orders of magnitude longer, albeit the memory efficiency is slightly lower. We attribute the limited lifetime to two reasons. One is the atomic loss from the optical lattice during storage. The other is imperfect magnetic compensation of the differential light shifts. Since the angles among the lattice beams are not zero, therefore the degree of circular polarization $A$ ($A=\pm 1$ for pure $\sigma^{\pm}$ light)~\cite{derevianko2010theory} of the lattices after interference is not constant for the whole atomic ensemble. Another reason that may cause imperfect magnetic compensation is the hyperpolarizability~\cite{yang2016coherence}.

After that, we test correlations of our atom-photon entanglement and its temporal dependence. Measurement is performed via measuring the polarization correlations between the write-out photon and the read-out photon both in the eigen basis $\ket{H}/\ket{V}$ and in the superpositional basis $\ket{D}/\ket{A}$, where $\ket{D}=1/\sqrt{2}(\ket{H}+\ket{V})$ and $\ket{A}=1/\sqrt{2}(\ket{H}-\ket{V})$. The visibility $V$ is defined as
\begin{equation}
V_{ab}=\left| \frac{n_{ab}+n_{ba}-n_{aa}-n_{bb}}{n_{ab}+n_{ba}+n_{aa}+n_{bb}}\right|
\end{equation}
where $n_{ab}$ is number of coincidences measured for a write-out photon in polarization $\ket{a}$ and a read-out photon in polarization $\ket{b}$. The temporal dependence for $V_{HV}$ and $V_{DA}$ is shown in Fig.~\ref{fig:lifetime}(b). We find that $V_{HV}$ and $V_{DA}$ nearly stay as constant values for $t\leq1$~s. For $t>1$~s, the visibilities start to drop significantly due to the reduction of signal-to-noise ratio. We also note that $V_{DA}$ is slightly smaller than $V_{HV}$, which is due to imperfect phase control of the interferometer. By selecting the points $t\leq1$~s, we get an average value of 0.935 for $V_{HV}$ and 0.879 for $V_{DA}$, which together enable us to estimate an entanglement fidelity as $F\simeq(1+\overline{V}_{HV}+2\overline{V}_{DA})/4$ = 0.923~\cite{Guhne2009}.

\begin{table}[htb]
\caption{Measurement of Bell inequality.}
\begin{tabular*}{\columnwidth}{@{\extracolsep\fill}ccccc}\toprule
$t$ & 5~$\upmu$s & 200~ms & 500~ms & 1~s \\
\hline
S value &2.64(9)&2.59(11) &2.41(12) &2.36(14) \\
\toprule\end{tabular*}
\label{tab:CHSH}
\end{table}

Finally, we verify the atom-photon entanglement directly via testing the Bell inequality~\cite{bell_problem_1966,clauser_proposed_1969}. We measure the S parameter for several different storage durations, with results shown in Tab.~\ref{tab:CHSH}. It is clear that, all the S values measured are above the threshold of $S>2$ to justify entanglement. Most notably, after storage of 1 second, we still get an S value of $2.36\pm 0.14$, which violates Bell's inequality by 2.57 standard deviations. To the best of our knowledge, this is a Bell test with the longest storage duration. The delay is already enough to perform Bell test with random basis selection by the free will of human beings.

In summary, we have realized a source of atom-photon entanglement with cavity enhancement and 3D-lattice confinement. We harness the momentum vector degree for a collective atomic excitation and the polarization degree for a single photon that is very robust for long-distance transmission in optical fibers. Together with quantum frequency conversion from near-infrared to telecom, our entanglement will become a building block to construct heralded entanglement between two remote quantum nodes~\cite{yu_entanglement_2020}. The achieved long-lived storage and efficient retrieval will enable scalable extension to multiple nodes and longer distance via photonic entanglement swapping~\cite{Sangouard2011}.

This work was supported by National Key R\&D Program of China (No. 2017YFA0303902), Anhui Initiative in Quantum Information Technologies, National Natural Science Foundation of China, and the Chinese Academy of Sciences.

\clearpage

\setcounter{figure}{0}
\setcounter{table}{0}
\setcounter{equation}{0}

\onecolumngrid

\global\long\def\theequation{S\arabic{equation}}
\global\long\def\thefigure{S\arabic{figure}}
\renewcommand{\thetable}{S\arabic{table}}

\newcommand{\msection}[1]{\vspace{\baselineskip}{\centering \textbf{#1}\\}\vspace{0.5\baselineskip}}
\msection{SUPPLEMENTARY INFORMATION}

\section{Efficiency of single-mode retrieval}

We first consider the case of single mode retrieval. With an atomic ensemble in free space, the retrieval efficiency $R_{\text{eff}}$ satisfies the following formula~\cite{gorshkov2007photon1,gorshkov2007photon2},
\begin{equation}
R_{\text{eff}}=\frac{C}{C+1},
\end{equation}
where $C$ is a cooperative parameter related to optical depth. With help of a ring cavity, the parameter $C$ can be enhanced further, resulting in a single-mode retrieval efficiency of
\begin{equation}
R_\text{sg}=\frac{C_\mathcal{F} \cdot C}{C_\mathcal{F} \cdot C+1},
\label{Rsingle}
\end{equation}
where $C_\mathcal{F}$ is the cavity enhancement factor and takes a value of $2\mathcal{F}/\pi$~\cite{bao2012efficient, heller2020cold}. Therefore, the single-mode retrieval efficiency is intrinsically limited by the optical depth and cavity finesse $\mathcal{F}$. Technically, the single-mode retrieval efficiency is further limited by a number of issues, such as intra-cavity losses which reduce photon escaping rate out of the cavity, fluctuation of the cavity resonance, mode matching between the write and read beams, mode matching between cavity output and coupling fiber.

\section{Efficiency of double-mode retrieval}

The double-mode retrieval involves two read beams. One read beam will collectively retrieve the atomic spin-wave into the cavity mode, while the other read beam will collectively retrieve into a phase-matching direction other than cavity mode. In addition, there is spontaneous emission in random directions induced by the two read beams. Thus, we obtain a double-mode retrieval efficiency as
\begin{equation}
R_\text{db}=\frac{C_\mathcal{F} \cdot C}{C_\mathcal{F} \cdot C+C+2}.
\label{Rdouble}
\end{equation}
There are three terms in the denominator of Eq.~\ref{Rdouble}. The first term $C_\mathcal{F} \cdot C$ represents retrieved photons going along the cavity mode. The second term $C$ represents photons in the other collective enhanced direction. The third term refers to photons read out to random directions. Combining Eq.~\ref{Rsingle} and Eq.~\ref{Rdouble}, we obtain the relation between single-mode efficiency and double-mode efficiency as following,
\begin{equation}
R_\text{db}=1/\left( 1+\frac{\pi}{2\mathcal{F}}+2\cdot\frac{1-R_\text{sg}}{R_\text{sg}} \right).
\end{equation}
We plot the theoretical curve in Fig.~2(b) based on this equation.

\section{Experimental sequences}

\begin{figure}[htb]
	\includegraphics[width=0.8\columnwidth]{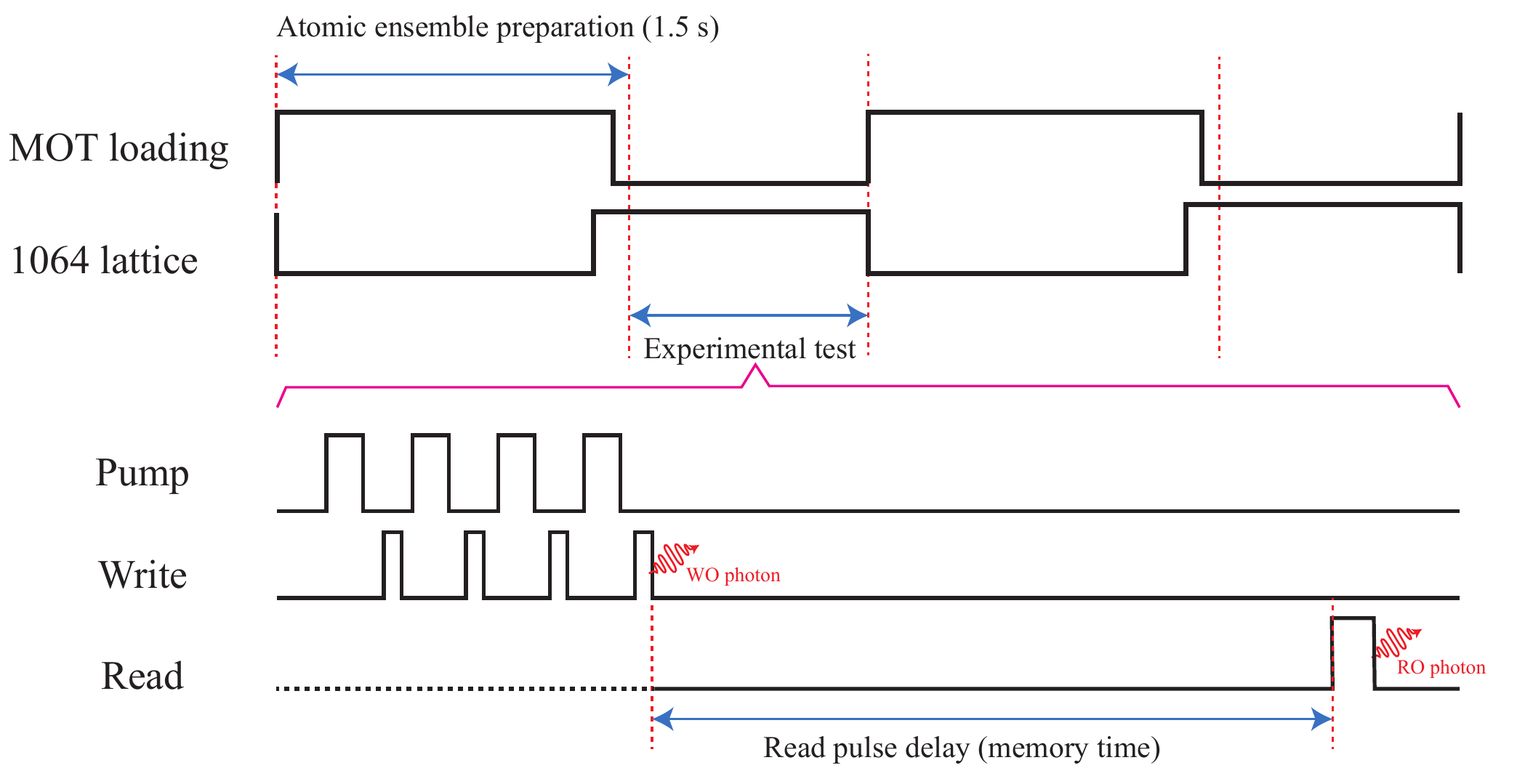}
	\caption{Experimental sequences.}
	\label{sequence}
\end{figure}
The experimental sequences are shown in Fig.~\ref{sequence}. Our experiment contains two phases. In the first phase we do atom loading, while in the second phase we perform experimental test. In the atom loading phase, we spend 1.5 second to prepare the atomic ensemble. Duration of the experimental test phase is not fixed, and it mainly depends on the storage time. In the experimental test phase, we apply pump and write pulses alternatively until a write-out (WO) photon is detected, where the pump pulse is used to pump atoms to one specific Zeeman sublevel. The detection of a white-out photon heralds a correlated atomic excitation that is stored in the ensemble. After a configurable time delay, the read pulse is applied to transform the atomic excitation to a read-out (RO) photon.

The write pulses are applied at most 400 times after each atom loading phase. The probability of detecting a write-out photon in each write process is 0.4\%. So after each atom loading, we can get a write-out photon with a high probability around 80\%. In average, 250 write pulses can generate one single write-out photon. The period of write pulse is 5 $\upmu s$, so the time needed in generating one pair of atom-photon entanglement is about 1.25 ms. If we take the atomic loading time into account, the entanglement production rate will be dominated by the atom loading time that can be further reduced significantly by optimization.

\end{document}